APPLIED CYBERSECURITY & INTERNET GOVERNANCE

ACIG

NASK# Digital Sovereignty Strategies for Every Nation

Received: 06.11.2022
Accepted: 14.11.2022
Published: 17.11.2022**Ali Shoker** Resilient Computing and Cybersecurity Center (RC3), Computer, Electrical and Mathematical Sciences and Engineering Division (CEMSE), King Abdullah University of Science and Technology (KAUST), ORCID: 0000-0002-4898-9394Cite this article as:
A.Shoker, "Digital Sovereignty Strategies for Every Nation,"
ACIG, vol. 1, no. 1, 2022.
DOI: 10.5604/01.3001.0016.0943

Corresponding author:
Ali Shoker, King Abdullah
University of Science and
Technology (KAUST), 23955-6900
Thuwal, Kingdom of Saudi Arabia;
ORCID: 0000-0002-4898-9394;
Email: ali.shoker@kaust.edu.saCopyright: Some rights reserved:
Publisher NASK. Publishing House
by Index Copernicus Sp. z o. o.## Abstract

Digital Sovereignty must be on the agenda of every modern nation. Digital technology is becoming part of our life details, from the vital essentials, like food and water management, to transcendence in the Metaverse and Space. Protecting these digital assets will, therefore, be inevitable for a modern country to live, excel and lead. Digital Sovereignty is a strategic necessity to protect these digital assets from the monopoly of friendly rational states, and the threats of unfriendly Malicious states and behaviors. In this work, we revisit the definition and scope of digital sovereignty through extending it to cover the entire value chain of using, owning, and producing digital assets. We emphasize the importance of protecting the operational resources, both raw materials and human expertise, in addition to research and innovation necessary to achieve sustainable sovereignty. We also show that digital sovereignty by autonomy is often impossible, and by mutual cooperation is not always sustainable. To this end, we propose implementing digital sovereignty using Nash Equilibrium, often studied in Game Theory, to govern the relation with Rational states. Finally, we propose a digital sovereignty agenda for different country's digital profiles, based on their status quo, priorities, and capabilities. We survey  state-of-the-art digital technology that is useful to make the current digital assets sovereign. Additionally, we propose a roadmap that aims to develop a sovereign digital nation, as close as possible to autonomy. Finally, we draw  attention to the need of more research to better understand and implement digital sovereignty from different perspectives: technological, economic, and geopolitical.

**Keywords**
autonomy, digital sovereignty, digital strategy, Nash Equilibrium, sovereign technology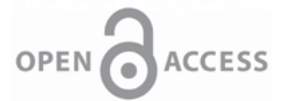

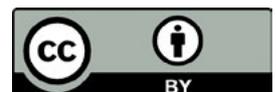

## 1. Introduction

Digital sovereignty is becoming a compelling priority to "control the present and the destiny" of modern nations [1], and a "make-or-break" issue [2]. It is a necessity for state's independence and national security in face of the increasing threats on *digital assets*. These threats are shown to be caused by both *like-minded* countries

www.acigjournal.com    ACIG, VOL.1, NO.1, 2022    DOI: 10.5604/01.3001.0016.0943    INDEX COPERNICUS INTERNATIONAL    1

[1] The primary goal of a Malicious adversary is to cause harm to the target; whereas the primary goal of a Rational adversary is to increase its own payoff (utility), regardless of the target. A Rational player is selfish, but not necessarily Malicious. See details in Section 3.

(e.g., USA, Germany, Canada, and Brazil) [3–7] and *non-like-minded* countries (e.g., USA, Russia, China, and Iran) [2, 8–10]. Protecting the digital assets from these threats, whether rational or malicious[1], is vital for modern countries given the unprecedented invasion of digital technology to our daily life essentials. The lack of digital sovereignty can undermine the automated and smart digital systems like water systems, food supply, smart power grids, telecommunications, Internet of Things, economy, health, governance, security, and defense [10–14]. Therefore, if these digital assets are threatened and compromised, national security will be at stake and human lives will be endangered.

The recent geopolitical tensions show that digital sovereignty is an urgent aspect more than ever. For instance, the recent bans of the USA on Huawei were mainly to protect the former's national digital sovereignty from undermining the telecommunication infrastructure and data [15, 16]. The tensions between the USA and China on the Taiwanese front are caused by the desire to control (around 70%) of the deep-tech semiconductor market fabrication [17]. The *Cyberwar* of state-backed malicious groups on websites in Finland, Italy, Romania, Germany, Norway, Lithuania, Czechia, Latvia, and elsewhere, is unprecedented [12]. Misinformation and systematic infiltration of social media can influence elections and democratic processes [2, 18]. Last, but not least, friendly relationships between countries cannot prohibit the surveillance of the officials of leading countries, like Germany [3].

Despite its importance, the study of Digital Sovereignty is still in its infancy. Since it touches upon different research areas like geopolitics, technology, and economics, more rigorous research efforts on the topic are still needed to fully understand the topic in a comprehensive and exhaustive way [18–20]. Digital sovereignty was originally promoted as data privacy and ownership - driven by political propagandas for "internal legitimacy" – to please the citizens [3, 18, 19, 21]. Then, it was defined and studied in different technological sovereignty areas, including digital, network, data, spectrum, Internet, cybersecurity, computer, and information [3, 22, 23]. Recently, the definition of digital sovereignty has got more attention in two dimensions. The first is on economic monopoly and intellectual property, that mainly targets the semiconductor fabrication, 5G infrastructure, and misuse of Artificial Intelligence (AI) with data [2, 18, 24–27]. The second dimension is related to the *Cyberwar* [14, 19, 28–30,]. Although these works emphasize important perspectives, they lack a definition that captures the big picture and, thus, leading to an incomplete perception of the issue, while leaving gaps in approaching it.

We therefore introduce a comprehensive definition of digital sovereignty that covers the entire value chain of a state's digital assets. Our definition (discussed in Section 2) captures the entire scope of the digital sovereignty's spanning data, infrastructure, fabrication, raw material, operational resources (raw material and humans), and research & innovation. The latter three aspects are noteworthy because they are often discarded or underestimated in literature. Several countries outsource their data and security operations (i.e., giving up digital sovereignty) to a handful worldwide known companies because of the lack of trained operational expertise and subject matter knowledge [1, 19, 31, 32]. China and Taiwan are leading the 5G and semiconductor manufacturing sectors because of their research advancements [15–17]. Similarly, the EU is lagging behind the USA and China in AI due to the lack of the highest *caliber* of talent and enough investment in R&I [32, 33]. On the other hand, reality shows that the shortage in energy supply as operational material can lead to a major digital shutdown, as in Ukraine and Lebanon [34–36]. We discuss the scope of digital sovereignty in detail, accompanied with a threat analysis to highlight the relevance and severity on these aspects.

The *means* to achieve digital sovereignty is even more challenging. We highlight the two main strategies proposed in literature and practice: sovereignty by *autonomy* and by *cooperation*. We show that although autonomy is the most effective sovereign way that states should seek, it is *impossible* with the current geopolitical landscape.

This gap is often bridged with the cooperative treaties and alliances, e.g., following the cooperative *bargaining* problem studied in *Decision Theory* and *Game Theory* [37, 38]. Nevertheless, we notice that the digital sovereignty problem incurs a notion of threat *par excellence*, which we address by suggesting another strategy called *Nash Equilibrium Sovereignty*.

*Nash Equilibrium Sovereignty* follows the *Nash Equilibrium* strategy in *Game Theory [7, 38], which* is a hybrid cooperative and non-cooperative strategy, in contrast to the cooperative *bargaining* strategy. Nash Equilibrium allows two players to converge to a stable situation without direct intentions to cooperate. Using Nash Equilibrium is reasonable since it is tailored, by definition [38], to games where some notion of threat is present, similar to the digital sovereignty *game* in our study. In particular, it targets the case of rational players, which well represents the modern state governance. This strategy should precede bargaining whenever possible, since it is more sustainable and guaranteed given the unexpected tensions that can arise between friend states or allies [3, 16, 40, 41]. For instance, the best strategy for two neighboring (although not-like-minded) states is to cooperate on Internet packet delivery. The best strategy for a *deep-tech* producer and a corresponding rare material producer is to exchange their production.

We argue that digital sovereignty should be on the agenda of every modern government that embraces the digital world. While this topic is indeed among the top priorities of some states, like the EU and USA [14, 32, 42–44], many non-developed or developing countries consider themselves unconcerned, either because of underestimating its impact, or considering it a *dream* – due to the lack of capacities to implement it. We alleviate these misconceptions by introducing a *digital sovereignty agenda for every nation*, considering three country profiles that represent the digital status quo of all countries: *User*, for countries that manly outsource data; *Owner*, for countries that purchase and own infrastructure; and *Producer*, for countries that manufacture or develop digital technology. Th producer level is the ultimate target of sovereign states because it leads to digital autonomy and reduces external dependencies. The proposed agenda stands as a high-level roadmap for governments to (1) ensure an attainable level of digital sovereignty defined according to its posture, and (2) lift its profile to the most ambitious level, i.e., the Producer. As a case study, we drive a non-exhaustive survey of the recent technological techniques and security countermeasures that can be used in implementing digital sovereignty. We then demonstrate which means, among these techniques, can be used by the three profiles to protect their prioritized digital assets.

Our conclusion draws attention to the lack of enough studies on understanding digital sovereignty, especially those that study the interplay between technology, economy, and geopolitics. We particularly encourage further research on the *means* used to achieve it.

## 2. A Comprehensive Definition and Scope of Digital Sovereignty

The recent interest in digital sovereignty led to many definitions in the three worlds: public, technology, and geopolitics [2, 3, 18–20, 24–27]. Unfortunately, none of those are sufficiently comprehensive due to their focal perspective, which may impede the comprehensive and exhaustive study & implementation of digital sovereignty. For instance, there has been a huge emphasis on data sovereignty, that often restricts digital sovereignty to data privacy and ownership, targeting citizen's legitimacy, i.e., pleasing the people [3, 18, 19, 45]. Many technological sovereignty definitions have considered the sovereignty of different technological fields like Technological sovereignty, Digital sovereignty, Network sovereignty, Data sovereignty, Spectrum sovereignty, Internet sovereignty, Cyber sovereignty, Computer sovereignty, Network sovereignty, and Information sovereignty [3, 22]. Geopolitical definitions got inspired by State Sovereignty [3, 46] and focused on selected digital facets in the realm of the Cyberwar or the monopoly of resources [14–16, 19, 28–30]. We bridge this gap by providing a comprehensive definition to digital sovereignty as follows:

### 2.1. Definition 1

*Digital Sovereignty* of a state is possessing the supreme authority over all its digital assets, including the entire value chain: data, infrastructure, operations, supply chain, and knowledge.

The above definition is derived from the definition of *State Sovereignty* [3, 46] and applied to the state's "digital assets". The salient novelty in our definition is considering the entire digital value chain of a digital asset. Although the most valuable digital asset is the "data" itself; it could be undermined if digital sovereignty does not address the entire value chain that generates, stores, processes, operates, and manages the data – as we explain next.

*State Sovereignty as an inspiration.* State Sovereignty is commonly defined as possessing the supreme authority over a territory [3, 46]. This authority is manifested as having full control over the territory. This definition should not, however, restrict the term "territory" to the landmass of a state; it rather includes the entire assets lying above and underneath, like the people, animals, space, air gases, oils, minerals, etc [47, 48]. In this sense, it is understood that the "territory" here represents the entire collection of assets that a state possesses (even if they reside abroad).

*Data, the core digital asset.* We argue that the entirety of the existence and importance digital assets are for the sake of "data", deemed here as the "core digital asset". Data is the digital[2] representation form of any piece of information. Data can have an immense impact on the state that grows as the reliance on digital technology grows; it is, therefore, inevitable for any modern and developing country. Data can be processed and presented as useful insights to make thoughtful and actionable decisions, or used to autonomously control other vital Cyber-Physical assets like power grids, water distribution, transportation, smart factories, etc. Unfortunately, experience shows that compromising these systems can endanger national security, human lives, and democracy [2, 11, 18].

*The "borders", or lack thereof.* Nevertheless, the "borders" in our definition to digital sovereignty are softer than those defined in State Sovereignty. As Barlow explained more than two decades ago, the "[c]yberspace does not lie within your borders" [49]. In fact, modern countries make an extensive use of the Internet and other digital communications within and outside their physical borders, e.g., for social, economic, military, and governance matters. Without these global communication channels, it is not difficult to figure out how slow and constrained the governance, life, and economy would be in this era. Nevertheless, data that is transported off-borders often uses other's communication channels, stored in remote storage, and processed using remote processors and software. These are all non-state-owned digital technology over which the data-owning state has little to no control. This is analogous – though more complex – to controlling other state assets oversees, e.g., ships and diplomatic representations in other countries. Likewise, data within borders can also be compromised by physical intruders, spies, thefts, and cyberattacks [8, 9, 13, 50].

### 2.2. Scope: the value chain, beyond data

Data is futile if not stored, processed, and transported. This is only possible through maintaining a large and complex value chain that is key in defining the correct scope of the digital sovereignty. We sketch this scope visually in Fig.1. The scope includes seven domains or aspects that cover the entire digital value chain. Data, infrastructure, fabrication, raw material, operational material, research & innovation (R&I), and operations. The latter two domains span the entire spectrum of the former five stacked domains. As explain earlier, the main emphasis of literary definitions was on the software and hardware stacks underlying the data, with partial focus on fabrication and material. Nevertheless, the operational material, R&I, and operations, have got little notice despite their essential role, compulsory to digital assets.

---

[2] Digital data is, typically, binary values of a physical quantity such as voltage or magnetic polarization.

We demonstrate the importance of the seven domains in Fig.1, highlighting the corresponding threat model for each. The threat model considers both the case of *rational and malicious threats. The former mainly addresses the economic monopoly of digital technology and intellectual property. The latter is studied in the light of the well-known CIA triad:* Confidentiality, Integrity, and Availability [51, 52]. In a nutshell, confidentiality specifies that unauthorised users cannot access or disclose the data, mainly due to privacy or intellectual property reasons. Integrity ensures that data at-rest or in-transit is genuine and consistent, i.e., not tampered with by unauthorized users. Availability ensures that data is always available to be read and/or updated by authorised users. There are several ways to violate these properties; some of which we discuss in the following points.

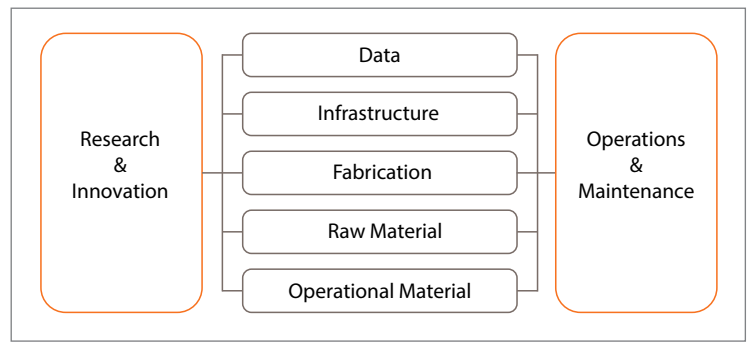

**Figure 1.** The architecture of the modern Internet, adapted from [4].

- **Data.** *Who rules the data, rules the world.* Data is the most valuable digital asset for which the entire value chain exists. It can hold sensitive state secrets, critical cyber-physical operational data, (bio)medical data, social data, and meta-data (i.e., data of data). Data is prone to threats on the three CIA triad [51]. Nevertheless, confidentiality can be easily compromised if it is outsourced for storage or computation outside the state's territory. Although a data operator may protect the data from unauthorized access, the data is under its mercy [1, 3]. This adversary (operator) is often called trusted-but-curious. This is why many countries have set regulations to protect outsourced data. Availability is mainly violated by *Denial of Service (DoS)* attacks or even encrypted for ransom reasons. For instance, the *WannaCry* ransomware attack has more than 230,000 devices [8, 50]. Unfortunately, many of these cyberattacks are believed to be coordinated by armed or hired groups by governments [8, 53].
- **Infrastructure (digital).** This is the set of software and hardware without which data cannot be: maintained, stored, processed, or transported. Digital infrastructure plays a main role in transforming data into useful information. It can be software such as: operating systems, protocols, AI algorithms, routing protocols, drivers, office tools, search engines, websites, etc. It can also be hardware that represents computers, servers, routers, cables, embedded devices, Integrated Circuits (ICs), etc. The application scope is also immense, spanning social networks, mobile applications, browsers, operating systems, propriety design, diagnostic tools, simulators, and testbeds. The threats on software and hardware can violate digital sovereignty in many ways. A malware or a backdoor in an imported software can compromise the three CIA properties [51]. In the last decade, we noticed more resistance to stop the hegemony of US-based software. There are continuous calls and attempts in the EU to find rivals to the USA's GAFAM (Google, Amazon, Facebook, Apple, Microsoft) and Chinese BAT (Baidu, Alibaba, and Tencent) companies [1, 26, 32, 54]. Until recently, the USA is also suffering from serious infrastructural threats. For instance, in one year, 649 critical infrastructure entities had been attacked according to the FBI [11]. *SolarWinds Orion Platform* has also been subject to one of the largest infrastructure's *backdoor* attacks in the USA [10]. These attacks can even be more critical, like the famous *Stuxnet* malware on the Iranian nuclear-fuel plant [13]. Finally, although the attack on the Russian *Yandex Taxi* remained mild and contained, it represents a simple example of what could happen if the semi-autonomous transportation infrastructure is attacked [12].

[7] For details see New Urban Communities Authority (newcities.gov.eg/english/New_Communities/badr/default.aspx).

- **Fabrication.** This is the most complex part in the value chain as it needs deep tech and rare raw materials. Hardware is rarely manufactured by one country and can be computers, mobiles, telecommunication equipment, datacentres, vehicles, medical equipment, smart devices, manufacturing control devices, modern vehicles, etc. In particular, silicon-hardware fabrication, e.g., semiconductors, requires decent technological advancements and industrial capabilities that exist in a handful of countries [16, 17]. This is critical from a digital sovereignty perspective since a producing country can ban any item in the supply chain, causing serious impacts on other countries [15, 16]. This is an economic and political threat of paramount significance. For instance, the sanctions on Russia are thought to cause a shortage in semiconductor chips that forced the Russian military to reuse chips from dishwashers and refrigerators [25]. *Huawei* lost its position at the top of mobile manufacturers due to bans by USA, Australia, New Zealand, Japan, and Taiwan [15, 16]. We also shed the light on the most serious attacks from a sovereignty perspective: trojans and backdoors [55, 56]. These are critical by being not easily detectable, and the producer is often supposed to be trusted. A state can buy a compromised hardware, e.g., military device, with a backdoor trojan that can be controlled by the producer/vendor [56]. A trojan can also play the "kill Switch" and, thus, activate itself after some time to ambiguate the attack for the buyer. The Israeli attack on the Syrian radar in 2007 is thought to be caused by a kill switch in an imported radar [56]. Similarly, it is extremely hard to verify if kill switches exist in other exported military weapons, like aircrafts. Nevertheless, these backdoors do exist in reality, like the one found in an Alcatel's military-grade field-programmable gate array (FPGA), although the company denies prior knowledge of it [57].
- **Raw material.** Fabrication requires a huge number of essential raw material, without which the product cannot be realized. The most typical example is semiconductor's resources that includes precious solids like *Gold, Aluminium, Diamond, Gallium Nitride and Silicon Carbide* [58–61]. These materials are used to produce the main hardware fabric primitives *Diodes, Transistors, Thyristors,* and other ICs [60, 62–64], using more than one hundred scarce and *Noble gases*, produced in very few countries [60, 65, 66]. The economy of these materials is subject to perturbations or shortage due to tensions between countries. For instance, the USA estimates that a Chinese invasion of Taiwan could disrupt the world economy and yield a loss of more than a trillion US dollars [17]. The recent Russian-Ukrainian war is not far from this monopoly, as Ukraine is a major producer of many noble gases – critical for semiconductor fabrication. The authors [67, 68] claim that if Russia retains control of *Mariupol* (a major Ukrainian factory where many Nobel gases like Neon, Krypton, Xenon, and Helium are produced) and restarts the city's damaged plants, 95% of the market could wind up in the hands of Russia and China.
- **Operational material.** These are raw materials needed for the operation of the infrastructure like energy and cooling sources, e.g., gas, oil, hydrogen, sunlight, wind, air, ice, etc. The entire value chain becomes useless if these operational resources are not guaranteed. For instance, the recent power outages due to the war in Ukraine and the Lebanese economic crisis led to major cut offs in the Internet and telecommunication industries [34–36]. On the other hand, Nordic countries are attracting dozens of major worldwide data centres due to the natural ice cooling [69]. This makes it more appealing and affordable to export data to these datacentres, but can also undermine data sovereignty as we explained above.

- **Research and Innovation (R&I).** This is a strategic aspect since human and institutional resources guarantee sustainable digital sovereignty through the quest for novel solutions and innovations [32, 70]. Reality shows that even leading countries can lose their position as a consequence of the lack of advancement in R&I. For instance, Taiwan's leadership in semiconductor fabrication is a result of consistent research and investment [32]. The Chinese leap in 5G and AI is referred to the higher percentages of R&D employees, as reported by Goldman Sachs [32]. The lag of the EU in AI innovation gives the impression that "China and the US innovate, while Europe can only regulate" [70].
- **Operations and Maintenance.** This includes the trained human resources with the subject-matter expertise to be able to operate the infrastructure. The lack of these trained resources can lead to sovereignty issues due to the need to outsource the data, or hiring third-party entities to maintain the operations [31]. Outsourcing sensitive data or cybersecurity data can compromise the countries' digital sovereignty since confidentiality and integrity would be violated as discussed earlier [26, 31]. Unfortunately, the worldwide regulations GDPR, CLOUD, among others [18, 21, 71, 72], fall short to mitigate these threats fully.

This wide scope and threat model indicates the complex nature of ensuring digital sovereignty at all levels. This represents a challenge in finding the correct *means* to ensure digital sovereignty. In addition, the above study shows that these means are not only technical, but rather economic, educational, and geopolitical by nature. This requires following economic strategies, geopolitical diplomacy and using special techniques and countermeasures to ensure data sovereignty, as we show next.

## 3. Strategies for Digital Sovereignty

The means to address digital sovereignty is cumbersome given its wide scope and the threats discussed above. The recent surveillance, leaks, threats, and incidents [3–6, 8, 9, 18, 50] called for compelling research to study the approaches towards sovereignty [16, 18, 22, 32, 43, 44, 46, 54, 73]. Two main directions are being heavily investigated: (1) autonomy that tries to build on self-reliance and independence to reduce the extremal influences [16, 18, 43], and (2) partnership that builds on cooperative bargaining through allies and coalitions to bridge the individual state gaps [16, 18, 28]. We show here that although both methods are useful, they fall short to achieve the sought digital sovereignty completely. Therefore, we propose using a complementary way that is more effective and reliable for *rational* behaviours, like countries. This approach is a hybrid cooperative and on-cooperative Game Theoretic model using *Nash Equilibrium*, introduced for rational behaviours by design [39, 49, 74].

### 3.1. Sovereignty by Autonomy

This approach promotes the independence and self-reliance means to defend against external threats [2, 16, 18, 43]. This should be the primary strategy used at the entire scope. While some purely technical domains like data, infrastructure, and operations are possible using state-of-the-art technology (as discussed later), other domains like fabrication and raw material seem more challenging. Although theoretically sound, this approach is often impractical for these domains with the current geopolitical landscape. Our conjecture is that digital autonomy is especially impossible in the case of small states. The reason is that the wide scope of digital sovereignty makes it very unlikely to autonomously supply and maintain the entire supply chain. Practice shows that even the digital sovereignty of leading and large countries, like the USA and China, can be undermined by the dependency on smaller or developing countries, like Taiwan [17]. While these large states can cope with it, this approach is undesired since it conflicts with other interesting properties like resilience [14, 18, 75], that embraces vendor diversity, economic openness to global consumers, and leadership that explores the best research ideas and minds worldwide [2, 24, 76]. It is noteworthy to mention that this approach may eventually lead to isolation and conservative relations with peer states, which can have a negative impact on the global and national technological advancements and national leadership. Note that the notion of "strategic autonomy" [18] is a little different from the autonomy we use here, as it may also refer to autonomy in decision making as well, which can itself follow several approaches, e.g., autonomy and mutual cooperation.

### 3.2. Sovereignty by Bargaining (i.e., cooperation)

This approach is complementary to the digital sovereignty by autonomy. It is based on building necessary partnerships with other states to bridge a gap in the national digital value chain [16, 18, 28]. Bargaining takes a form of unilateral treaties between states, that leads to mutual benefits, or multilateralism form where coalitions and alliances are built for the common benefit of the group, e.g., the EU, NATO, or Gulf Cooperation Council (GCC).

Technically, this approach is realized by solving the cooperative *bargaining* problem, studied in *Decision Theory* and *Game Theory* [37, 38]. Classical bargaining is based on negotiations where a decision *utility* (i.e., representing payoffs) of expected values in the future is maximized, as suggested in the *Neumann–Morgenstern (VNM) utility* theorem in 1953 [77]. More formally, a two-person bargaining problem consists of a pair *(F, d)*, where *F* represents the set of feasible agreements and *d is* the disagreement point (payoffs) if bargain terminates without an agreement. The solution is to find a function *f* that takes a bargaining problem *(F, d)* as input and returns a feasible agreement as outcome, i.e., *f (F, d)* ∈ *F*. Finding the solution *f* can follow different criteria, like maximizing the product of gains *(i.e., Nash Product)* [37, 38], equalizing the gains [49], among others [67, 74].

Nevertheless, bargaining has been theoretically criticized for being unrealistic for several reasons [39, 47, 74]. The most relevant reasons in our context are (1) assuming a coalition of all states, e.g., relevant to a digital domain, will form; and (2) it ignores the effects of external actions (from other states) to the coalition. In the digital world, these assumptions are unrealistic given the wide dimension, and the evolving nature of digital technology that cannot be defined or restricted. For instance, despite the dominance of the USA in software, hardware, and telecommunications, the Chinese TikTok and Huawei's 5G have made a breakthrough in these domains [15, 16, 32, 33]. On the other hand, extending the coalition by joining new relevant states is not always successful as per the current geopolitical landscape. For instance, the attempts to strengthen cooperation with Taiwan is recently yielding heated dispute between the USA and China [17]. In the EU, the membership of states with digital manufacturing capabilities, like Turkey; or Nobel gases production, like Ukraine, are witnessing resistance [78]. The same holds for the NATO's membership of Sweden and Finland [79]. The *EU's Brexit* is another typical example – not only at the digital front though – showing that even an existing membership may break [16]. Finally, there is an increasing Chinese bilateral "third country" [16, 40] influence on *central* and *eastern Europe* (e.g., Italy, Hungary, Slovenia, and Greece). This undermined the EU's unity in the realm of the Chinese *16+1 initiative, Belt and Road,* and *China-CEEC* [40, 41].

### 3.3. Sovereignty by Nash Equilibrium

We propose using the *Nash Equilibrium* strategy studied in *Game Theory* as a hybrid cooperative and non-cooperative strategy to digital sovereignty when autonomy is not viable [38, 39, 49, 74]. Our inspiration is referred to the nature of the digital sovereignty problem that, by definition, incurs a notion of threat between *rational* players. State governance is rational by excellence and the digital assets are highly prone to several threats, as described in the previous section. Our observation is that digital sovereignty by bargaining is not always realistic since it becomes a "solution applied to a wrong model", where states are assumed to be *benign* or even *altruistic*. This observation is consistent with the experimental studies on various bargaining models [47] showing that bargainers are found to focus on conceptually easy solutions that are beneficial to both parties. Nevertheless, bargaining is still a useful tool when sovereignty by Nash Equilibrium is infeasible.

Technically, a chosen strategy (of a set of actions) among possible ones is a Nash Equilibrium if no player can do better by unilaterally changing its strategy. More formally, let two players (i.e., states in our case) $A$ and $B$ have $S_A$ and $S_B$ as sets of possible strategies with utility functions (payoffs) $u_A$ and $u_B$, respectively. A binary setting ($s^i_A$, $s^j_B$), where $s^i_A$ $S_A$ and $s^j_B$ $S_B$, is a Nash Equilibrium if $A$ cannot obtain a higher utility payoff ($u_A$) than choosing $s^i_A$, i.e., $u_A(s^i_A, s^jB) > u_A(s^i_A, s^x_B)$ for any $sxB \in SB$. The same holds by symmetry for player $B$, with respect to strategy $sjB$ and utility $uB$. Since both $A$ and $B$ cannot do better, the game will stabilize, and the chosen strategies are enforced as if there is a cooperative agreement.

Practical examples on this approach are the cooperation of two *not-like-minded* neighboring states on Internet packet delivery. Nash Equilibrium is achieved since both states will deliver packets to their destination otherwise their own packets will be at stake. Another example in the semiconductor market is between the USA, that dominates the semiconductor design market [16], and Taiwan, that own the cutting edge 3*nm* and 5*nm* semiconductor fabrication [17]. Since available alternatives are scarce on both sides, none can efficiently produce semiconductors alone, which forces them to the Nash Equilibrium strategy. On the other hand, the wave of decentralized systems, inspired by *Blockchains*, are leading several use-cases in Fintech, Supply Chain, Cloud Computing, Governance, etc., that partially follow this rational model [80].

Sovereignty by Nash Equilibrium exhibits some drawbacks rooted in their design and application. Two drawbacks are particularly more relevant in our context. The first is the existence of multiple Nash Equilibria in one game, which sometimes prevents reaching the highest utility possible (i.e., called *Pareto optimality*) [39, 74]. The approach also assumes the knowledge of all states to all potential strategies, which might be too expensive or infeasible for small states. A *dummy* player who cannot attain this information may not be *rationally* on par with its counterpart, which violates the original assumption [39, 81].

Finally, we argue that the three strategies should be used together to ensure digital sovereignty. The preferred one must always be autonomy, followed by Nash Equilibrium. We recommend the latter over bargaining since it is more sustainable as discussed above. This is due to the intrinsic needs of the counterparts and, thus, imposing the Nash Equilibrium strategy can be seen as a *soft enforcement*. Bargaining can be an alternative to Nash Equilibrium when the latter is infeasible. Nevertheless, the cooperation between states is always encouraged in general, although it is ineffective when a rational player is not *playing fair*.

### 4. An Agenda for Digital Sovereignty

The above challenges indicate an unprecedented need for a digital sovereignty strategy for every nation. Nevertheless, it seems this represents a major concern for a limited number of countries [5, 15, 28, 32, 43]. The reason could be referred to underestimating the criticality of digital sovereignty or the lack of awareness. We have discussed earlier the criticality for the topic to every country embracing the digital world. In addition, non-developed countries may consider themselves unconcerned or not ready because of their limited capability to do anything about digital sovereignty.

To alleviate these misconceptions, we propose an agenda for the digital sovereignty targeting three profiles of nations, based on their capabilities and digital maturity. These profiles represent the majority of worldwide countries. The agenda includes an implementation of a state's current digital sovereignty as well as a future development plan. Being strategic, digital sovereignty follows a long duration roadmap that makes it an urgent priority for any modern government, sooner not later. We divide the agenda into three parts: self-assessment, planning, and implementation.

### 4.1. Self-assessment

1. *Establish a National Digital Sovereignty Agency (NDSA)*: this agency is in charge of the assessment, planning, implementation, and evaluation plans for digital sovereignty. It mediates the discussions with all other relevant agencies and ministries to digital assets. In particular, it can work under the supervision of a national state sovereignty agency if exists, and coordinates with other sovereignty agencies, e.g., food and borders sovereignty agencies.
2. *Identify national digital assets*: NDSA appoints relevant teams, workshops, and discussions to identify the state's digital assets.
3. *Drive a sovereignty threat and risk assessment*: all identified digital assets should be subject to a threat and risk assessment, covering the entire value chain discussed in the previous section.
4. *Designate the states' posture*: User, Owner, Producer. This gives the state a digital profile based on its status quo and capabilities. A posture of each digital asset can be assigned (more details below):
   - *User*: mostly uses digital technology and infrastructure that others produce or own.
   - *Owner*: mostly owns digital technology and infrastructure that others produce.
   - *Producer*: mostly produce all used technology and infrastructure.

### 4.2. Planning

A plan for *short-, mid-, and long-term stages should be defined*. Digital sovereignty is a long-road project that clearly requires a *long-term* roadmap (e.g., 30 years, for a developing country). However, it should be developed incrementally with *best effort* over a *mid-term plan* (e.g., 10 years), and a *short-term plan* (e.g., 3 years). The mindset and goal are to *advance the posture from User to Owner or Producer.* We do not envision technical restrictions that prohibit the planning from User to Producer postures in some aspects, although it may be infeasible in others. The proposed work-flow for the planning is as follows:

1. *Set goals based on the posture expected at each stage*: different digital assets may have different maturity levels (User, Owner, or Producer). This requires setting a national digital sovereignty goal for all digital assets at every stage. The purpose is to lift the digital maturity to a standard norm, preparing for the next stage.
2. *Set sector priorities based on its severity level*: it might not be feasible to focus on all sectors at once. A state may start with the most critical or sensitive sectors or assets.
3. *Adopt a sovereignty strategy*: the ultimate goal for nations is to achieve autonomy in all digital assets and sovereignty domains as in Fig. 1. This may not be achievable in some economies depending on digital maturity and geopolitical reasons. However, the goals should be set high enough for each state. In parallel, digital assets that rely on external Producers or Owners may follow a Nash Equilibrium strategy at first. As discussed in the previous section, this is because it is a more sustainable and reliable strategy for rational states. The rest of external dependencies can be sought through bargaining, e.g., building economic partnerships and alliances via diplomacy. Therefore, the three strategies should be used simultaneously.

### 4.3. Implementation

*Define and enforce legislations and regulations*: this is the very first actionable step that is needed to set the standards that national and international stakeholders should abide to. This should account for some time (e.g., few years) before enforcement, leaving a suitable window of time for stakeholders to prepare for a successful and smooth change.

*Apply State-of-the-Art (SotA) techniques and countermeasures*: this is the most technical part of the implementation. It makes use of SotA tools and techniques tailored to make digital assets sovereign, e.g., like Privacy, (Cyber)Security, and Resilience. These techniques should be researched and developed on a regular basis, otherwise some of them will be deprecated with time and fail against an evolving stronger adversary. For instance, security techniques that rely on classical cryptography, e.g., Public-Key Cryptography (PKC), may at some point be replaced with *post-Quantum* cryptography [82]. A high-level workflow would be as shown in Fig. 2.

### 4.4. Implementation case study

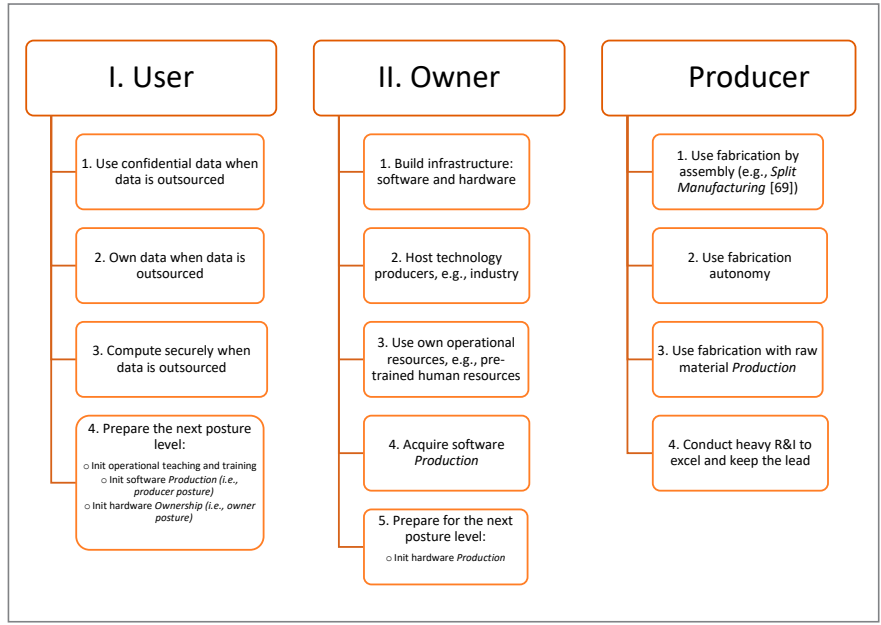

**Figure 2.** A high-level workflow for the implementation of digital sovereignty. The natural development plan is to acquire a User, then an Owner, then a Producer profile.

We present a case study to explain the methodology of the above implementation, showing that countries with different digital postures can do much on the sovereignty front. We demonstrate this in Fig. 3. and Tab. 1. The former conveys an example of selected state-of-the-art (SotA) techniques and countermeasures used for digital sovereignty. Although these techniques are part of SotA, the list is only meant to exemplify the state's capabilities and application feasibility – thus, it is not an exhaustive list. The right column represents five high-level techniques numbered from 1 to 5, mostly, but not strictly, in increasing level of complexity.

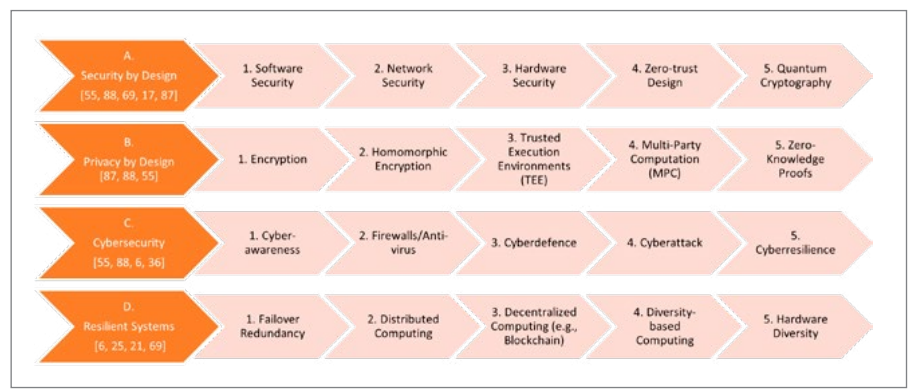

**Figure 3.** An example of state-of-the-art digital sovereignty techniques and countermeasures.

On the other hand, Tab. 1. matches the severity of applications or systems (in the left column) to the state's *posture* (in the bottom column). To be generic enough, we inspire the application severity from the "human needs" model in *Maslow's pyramid* [83, 84], with some exceptions for sensitive and deep technologies. However, a real-world case is to go over the entire digital-based services a government provide. Finally, we try to map the techniques used to the three defined postures. The main observation is that even states with (the lowest) User profile can improve their sovereignty with the currently available technology. More advanced postures, like the Owner and Producer have more capabilities to apply more techniques. We leave the details of this matching as an exercise to the reader, and we encourage a deeper study on this front.

**Table 1.** An example of matching the sovereignty techniques and countermeasures in Table 1 to a state's posture and severity levels (inspired by Maslow's pyramid of human needs [83, 84]).

|  |  |  |  |  |
|---|---|---|---|---|
| **Happiness** | Entertainment<br>Gaming<br>Art<br>Lesure | A.1-2<br>B.1<br>C.1-2 | A.1-2<br>B.1<br>C.1-3<br>D.1-2 | A.1-4<br>B.1-3<br>C.1-3<br>D.1-3 |
| **Needed** | Enivornment<br>Politics<br>Education<br>Social nets<br>Associations | A.1-2<br>B.1-2<br>C.1-2 | A.1-4<br>B.1-3<br>C.1-3<br>D.1-3 | A.1-5<br>B.1-5<br>C.1-5<br>D.1-4 |
| **Critical** | Air, Water, & Energy<br>Security, Top Secret<br>Deep tech.<br>(e.g., Space & Nuclear) | A.1-2<br>B.1-3,<br>C.1-2 | A.1-5<br>B.1-5<br>C.1-5<br>D.1-4 | A.1-5<br>B.1-5<br>C.1-5<br>D.1-5 |
|  |  | **User** | **Owner** | **Producer** |

## 5. Conclusions

Digital sovereignty is getting more traction due to the data surveillance, leaks, cyberattacks, and monopoly of digital resources. In this work, we have introduced a new definition to digital sovereignty showing that its scope must include the operational material, human resources, as well as research and innovation. Indeed, the lack of any of these aspects makes outsourcing data storage, computation, or operation a must, which undermines sovereignty. We have provided a threat model to emphasize the criticality of all the aspects of the digital scope, giving real-world examples. Then we proposed a new digital strategy by Nash Equilibrium to be used when autonomy is not feasible, while cooperative bargaining with other states should not be discarded. We also proposed an agenda for digital sovereignty to set the roadmap for a higher profile digital maturity. We show that any country can improve its sovereignty by following available techniques and countermeasures.

Our work can benefit from different directions in the future. First, a more exhaustive threat model and countermeasures are needed. Second, the bargaining and Nash Equilibrium strategies require deeper study to make the implementation of digital sovereignty easier. In particular, it is interesting to study how to play these games within coalitions (e.g., EU and GCC) and alliances (e.g., NATO) without breaking the member state sovereignties. Third, the agenda we propose is high level; a lower-level roadmap that stands as a detailed template for governments to follow would be very useful. Finally, the sovereign techniques and countermeasures we survey can benefit from a more comprehensive and thorough research in the future [85–88].

ACIG - Applied Cybersecurity & Internet Governance